\documentclass{article}
\usepackage{amsmath,amssymb,cite,epsfig,graphicx,fancyhdr}
\usepackage[latin1]{inputenc}
\begin{document}

\lhead{Factorization method and new potentials from the inverted oscillator}
\rhead{Bermudez, Fern\'andez C.}

\title{{\bf Factorization method and new potentials\\ from the inverted oscillator}}
\author{David Bermudez\footnote{{\it email:} dbermudez@fis.cinvestav.mx}\,  
and David J. Fern\'andez C.\footnote{{\it email:} david@fis.cinvestav.mx} \\
{\sl Departamento de F\'{\i}sica, Cinvestav, A.P. 14-740, 07000 M\'exico D.F., Mexico}}

\date{}

\maketitle

\begin{abstract}
In this article we will apply the first- and second-order supersymmetric quantum mechanics to obtain new exactly-solvable real potentials departing from the inverted oscillator potential. This system has some special properties; in particular, only very specific second-order transformations produce non-singular real potentials. It will be shown that these transformations turn out to be the so-called complex ones. Moreover, we will study the factorization method applied to the inverted oscillator and the algebraic structure of the new Hamiltonians.
\end{abstract}

\hspace{-5mm}{\bf Keywords} supersymmetric quantum mechanics; Darboux transformations; harmonic oscillator; inverted oscillator

\section{Introduction}
Let us consider the following Hamiltonian
\begin{equation}
H=-\frac{\hbar^2}{2m}\frac{\text{d}^2}{\text{d} x^2}+\frac{1}{2}m\omega^2 x^2,
\end{equation}
where $m$ has units of mass and $\omega$ of frequency. In order to simplify, we are going to use natural units, such that $\hbar=m=1$, to obtain
\begin{equation}
H=-\frac{1}{2}\frac{\text{d}^2}{\text{d} x^2}+\frac{1}{2}\omega^2 x^2.\label{Homega}
\end{equation}
Moreover, by choosing appropriately the value of $\omega$, three essentially different cases can be obtained:
\begin{equation}
\omega=
\begin{cases}
1 & \text{harmonic oscillator},\\
0 & \text{free particle},\\
i & \text{inverted oscillator}.
\end{cases}
\end{equation}
These are three rare examples of exactly-solvable potentials in quantum mechanics. The first one, the harmonic oscillator, is a very well known system from which the technique of creation and annihilation operators and the whole formalism of the factorization method come from. The second, the free particle, has also been largely studied. This simple system allows us to work close to the limits of quantum theory, for example, with non-square-integrable wavefunctions with plenty of physical applications such as the plane waves. The third case is not so familiar: it is called either \textit{inverted oscillator}, \textit{repulsive oscillator}, \textit{inverse oscillator}, or \textit{parabolic potential barrier}. Although it started as an exercise from Landau's book \cite{LL58}, its physical applications have grown since the appearance of Barton's PhD thesis (published in \cite{Bar86}), v.g., as an instability model, as a mapping of the 2D string theory \cite{YKC06}, or as a toy model to study early time evolution in inflationary models \cite{GP91}.

It is interesting to observe that both oscillator potentials, harmonic and inverted, are simultaneously produced inside an ideal Penning trap, typically used to confine charged particles \cite{bg86,fv09}. In its standard setup, a quadrupolar electrostatic field creates a harmonic oscillator potential along the symmetry axis of the trap, inducing confinement along this direction. In addition, in the orthogonal plane a two-dimensional inverted oscillator arises, driving the particles towards the trap walls. In order to compensate for the last effect, a static homogeneous magnetic field along the symmetry axis of the trap is also applied, but for zero magnetic field the two kinds of oscillator potentials are created inside the cavity.

Mathematically, the harmonic and inverted oscillators are very much alike, and we will show that the solutions of one can be obtained almost directly from the other one; nevertheless, we should remark that physically these two systems are very different. For example the harmonic oscillator has a discrete non-degenerate equidistant energy spectrum with square-integrable eigenfunctions, while the inverted oscillator has a continuous spectrum varying from $-\infty$ to $+\infty$, which is double degenerate, and whose eigenfunctions are not square-integrable.

On the other hand, a standard technique for generating new exactly-solvable potentials departing from a given initial one is the supersymmetric quantum mechanics (SUSY QM) (for recent reviews see \cite{mr04,ac04,su05,ff05,Fer10,ai12}). Its simplest version, which makes use of differential intertwining operators of first-order, has been employed for generating Hamiltonians whose spectra differ from the initial one in the ground state energy level. In addition, the higher-order variants, which involve differential intertwining operators of orders larger than one \cite{AIS93,aicd95,bs97,fe97,ast01}, allow as well the modification of one or several excited state levels.

The SUSY techniques of first- and higher-order have been successfully applied to the harmonic oscillator \cite{fh99,bf11} and the free particle \cite{ms91,bff12} for generating plenty of exactly-solvable potentials. However, as far as we know, neither the first- nor the higher-order SUSY QM have been employed taking as a point of departure the inverted oscillator. In this paper we aim to fill the gap by applying the supersymmetric transformations to the inverted oscillator. In order to do that, in Section 2 we will get the general solution of the stationary Schr\"odinger equation (SSE) for the Hamiltonian \eqref{Homega} with an arbitrary energy $E$, which will remain valid even for $E\in{\mathbb C}$. In addition, the solutions which have a physical interpretation for the inverted oscillator will be identified. In Section 3 we are going to explore the factorization method for both systems, obtaining the bound states for the harmonic oscillator and also several sets of mathematical polynomial solutions, a class of solutions which have been of interest along the time (see e.g. \cite{qu08,os09,gkm10}). In Section 4 we will work out the first-order SUSY QM for the inverted oscillator, while in Section 5 we will apply the second-order one in three different situations: real, confluent and complex cases. The last one will be the most important case of the paper, as it is the only one that works out to obtain new real non-singular potentials. In Section 6 we will explore the algebraic structure for the non-singular potentials generated through SUSY QM and their associated eigenfunctions. Finally, in Section 7 we will present our conclusions. The Appendix contains the derivation of the orthogonality and completeness relationships for the set of eigenfunctions of the inverted oscillator Hamiltonian (see also \cite{wo79}). 

\section{General solution of the stationary Schr\"odinger equation}

First of all, let us solve the stationary Schr\"odinger equation for the Hamiltonian \eqref{Homega} with an arbitrary real energy $E$, although it is still valid for $E\in{\mathbb C}$:
\begin{align} \label{sseq}
H\psi(x) = \left(-\frac{1}{2}\frac{\text{d}^2}{\text{d} x^2}+\frac{\omega^2 x^2}{2}\right)\psi(x)  = E \psi(x).
\end{align}
The substitution $x=\omega^{-1/2}y$ with $\phi(y) \equiv \psi(\omega^{-1/2}y)$ leads to
\begin{equation}
\left(-\frac{1}{2}\frac{\text{d}^2}{\text{d} y^2}+\frac{y^2}{2}\right)\phi(y) = \frac{E}{\omega} \, \phi(y),
\end{equation}
which is the SSE for the harmonic oscillator potential in the variable $y$ associated with the `energy'
$E/\omega$ \cite{JR98}. Here we assume that $\omega\neq 0$; otherwise we immediately get the free particle problem. Thus, the general solution to the SSE \eqref{sseq} reads
\begin{equation}
\psi(x)= e^{-\omega x^2/2}
\left[C {}_1F_1\left(\frac{1}{4} - \frac{E}{2\omega},
\frac{1}{2}; \omega x^2 \right) + D \, x \, {}_1F_1\left(\frac{3}{4} - \frac{E}{2\omega},\frac{3}{2}; \omega x^2\right)\right],\label{roy1}
\end{equation}
where ${}_1F_1$ is the confluent hypergeometric function and $C,D\in\mathbb{R}$ are constants. Note that, in general $\psi(x)\notin \mathcal{L}^2(\mathbb{R})$, i.e., it does not belong to the space of square-integrable wavefunctions in one dimension for an arbitrary $E\in\mathbb{C}$.

The analysis of the solution for the harmonic oscillator (take $\omega = 1$ in Eq.~\eqref{roy1}) is widely known and can be found for example in \cite{ff05,JR98}. On the other hand, since the inverted oscillator is not commonly studied, we will work it out next in detail.

The general solution of the inverted oscillator is obtained from Eq.~\eqref{roy1} for $\omega=i$. For simplicity, it will be expressed in terms of solutions with a definite parity, the even ($\psi_e$) and odd ($\psi_o$) solutions, which are given by
\begin{align}
\psi_e(x) & = \text{e}^{-ix^2/2} {}_1F_1\left(\frac{1+2iE}{4},\frac{1}{2};ix^2\right) = 
\text{e}^{ix^2/2} {}_1F_1\left(\frac{1-2iE}{4},\frac{1}{2};-ix^2\right), \label{ue}\\
\psi_o(x) & = x \, \text{e}^{-ix^2/2} {}_1F_1\left(\frac{3+2iE}{4},\frac{3}{2};ix^2\right) = x \, \text{e}^{ix^2/2} {}_1F_1\left(\frac{3-2iE}{4},\frac{3}{2};-ix^2\right). \label{uo}
\end{align}
This decomposition will simplify our mathematical work in Sections 4 and 5. Therefore the general solution is the following linear combination
\begin{equation}
\psi(x)=C \psi_e(x)+D \psi_o(x).\label{roy2}
\end{equation}
It is clear now that $\psi(x)$ is a real function, $\overline{\psi}(x)=\psi(x)$, for
any $E,C,D\in{\mathbb R}$.

Next, let us analyze the leading asymptotic behaviour of the solutions given in Eqs.~(\ref{ue}--\ref{uo}). To do this, we need the asymptotic behaviour of ${}_1F_1$ for $|z|\gg 1$ \cite{AS72}
\begin{equation}
{}_1F_1(a,b,z)\simeq\frac{\Gamma(b)}{\Gamma(b-a)}\text{e}^{i \pi a}z^{-a}+\frac{\Gamma(b)}{\Gamma(a)}\text{e}^{z}z^{a-b}. \label{limit}
\end{equation}
Hence, the asymptotic behaviour for $\psi_e$ and $\psi_o$ can be straightforwardly obtained for $x\gg 1$:
\begin{equation}
\psi_e(x) \simeq
\frac{\pi^{1/2} e^{-\pi E/4}}{x^{1/2}} \left[
\frac{e^{i(\pi/8 - x^2/2 )}x^{- i E}}{\Gamma(1/4-i E/2)} +
\frac{e^{-i(\pi/8 - x^2/2)}x^{iE}}{\Gamma(1/4 + i E /2)}
\right],\label{limue}
\end{equation}
\begin{equation}
\psi_o(x) \simeq
\frac{\pi^{1/2} e^{-\pi E/4}}{2x^{1/2}} \left[
\frac{e^{i(3\pi/8 - x^2/2)}x^{-iE}}{\Gamma(3/4-iE/2)} +
\frac{e^{-i(3\pi/8 - x^2/2)}x^{iE}}{\Gamma(3/4 + iE/2)}
\right].\label{limuo}
\end{equation}

By taking into account these equations, a complex linear combination is found such that now the terms going as $x^{-iE}$ get cancelled:
\begin{equation}
\psi_E^+(x) = N_E \left[\psi_e(x)- \frac{2\text{e}^{-i\pi /4} \, \Gamma(3/4 - i E /2)}{\Gamma(1/4 - i E/2)}  \psi_o(x)\right],
\label{uplus}
\end{equation}
where the `normalization' factor $N_E$ will be chosen in order to form an orthonormal set of eigenfunctions in the Dirac sense (see Appendix and \cite{wo79}). On the other hand, another linearly independent solution can be found from $\psi_E^+(x)$ through the reflection $x\rightarrow -x$:
\begin{equation}
\psi_E^-(x) = N_E \left[\psi_e(x)+ \frac{2\text{e}^{-i\pi /4} \, \Gamma(3/4 - i E /2)}{\Gamma(1/4 - i E/2)}  \psi_o(x)\right].
\label{uminus}
\end{equation}

It is worth noting that $\{\psi_E^\sigma(x), \ \sigma = \pm, \ -\infty<E<\infty\}$ is a complete orthonormal set of eigenfunctions for the inverted oscillator Hamiltonian satisfying the following orthogonality and completeness relations:
\begin{eqnarray}
&& (\psi_E^\sigma,\psi_{E'}^{\sigma'}) = \int_{-\infty}^{\infty} \overline{\psi}_E^{\, \sigma}(x)\psi_{E'}^{\sigma'}(x) dx 
= \delta_{\sigma,\sigma'} \delta(E-E'), \\
&& \sum_{\sigma = \pm} \int_{-\infty}^{\infty} dE \psi_E^\sigma(x) \overline\psi_E^{\, \sigma}(x') =  \delta(x-x'),
\end{eqnarray}
where $\delta_{\sigma,\sigma'}$ and $\delta(y-y')$ denote the Kronecker and Dirac delta functions respectively
(for a derivation of these equations we refer the reader to the Appendix and \cite{wo79}).

One can construct now a real linear combination with a specific physical interpretation \cite{MRW09,Wol10}, namely,
\begin{equation}
\psi_L(x) = \psi_e(x)- \left[ \frac{\text{e}^{i\pi /4} \, \Gamma(3/4 + i E /2)}{\Gamma(1/4 + i E/2)} +
\frac{\text{e}^{-i\pi /4} \, \Gamma(3/4 - i E /2)}{\Gamma(1/4 - i E/2)} \right] \psi_o(x). \label{ul}
\end{equation}
The subscript $L$ is employed due to $\psi_L(x)$ represents a particle incident from the left since its probability amplitude for $x<0$ is substantially bigger than the one for $x>0$ when $E<0$. On the other hand, another linearly independent real eigenfunction for the same $E$ with physical meaning can be obtained
from $\psi_L(x)$ through the reflection $x \rightarrow -x$:
\begin{equation}
\psi_R(x) = \psi_L(-x) = \psi_e(x)+ \left[ \frac{\text{e}^{i\pi /4} \, \Gamma(3/4 + i E /2)}{\Gamma(1/4 + i E/2)} + \frac{\text{e}^{-i\pi /4} \, \Gamma(3/4 - i E /2)}{\Gamma(1/4 - i E/2)} \right] \psi_o(x). \label{ur}
\end{equation}
The subscript $R$ denotes the fact that $\psi_R(x)$ represents now a particle incident from the right.

\begin{figure}
\begin{center}
\includegraphics[scale=0.65]{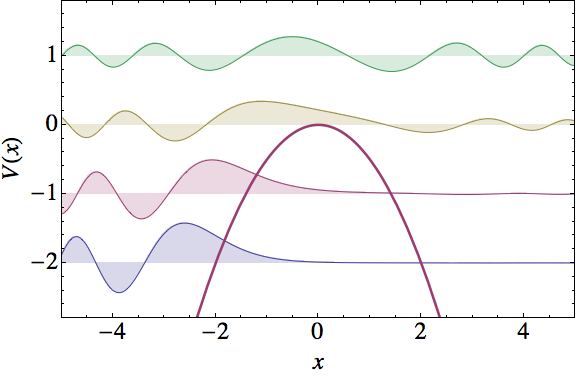}
\end{center}
\vspace{-5mm}
\caption{Eigenfunctions $\psi_L(x)$ of the inverted oscillator Hamiltonian given by Eq.~\eqref{ul} for the energies $E \in \{ -2,-1,0,1 \}$.} \label{figure1}
\end{figure}

The classical and quantum behaviour of a particle under the inverted oscillator potential is summarized in Table~\ref{table}. In Fig.~\ref{figure1} we have plotted as well the eigenfunction $\psi_L(x)$ for different values of the energy $E$, i.e., the quantum behaviour of a particle incident from the left, which illustrates the results of Table~\ref{table}. Since the complete set of orthonormal eigenfunctions of $H$ $\{\psi_E^\sigma(x), \ \sigma = \pm, \ -\infty<E<\infty\}$ has been found, one can conclude that the energy spectrum of the repulsive oscillator Hamiltonian consists of the full real line, each eigenvalue $E\in{\mathbb R}$ being doubly degenerated.

\begin{table}
\begin{center}
\begin{tabular}{cll}
\hline
Value of $E$& Classical behaviour & Quantum behaviour\\
\hline
$E >0$& Goes over & Most probable goes over, some is reflected\\
$E =0$& Is trapped an infinite time & Is trapped a finite time (sojourn time)\\
$E <0$& Is reflected & Most probable is reflected, some goes over\\
\hline
\end{tabular}
\end{center}
\vspace{-3mm}
\caption{The classical and quantum behaviour of a particle under the inverted oscillator potential.} \label{table}
\end{table}

\section{Factorization method}

The Hamiltonian \eqref{Homega} is hermitian for $\omega=1,0,i$ and it admits the well known factorization method. Indeed, by using the analogues of the annihilation and creation operators for the harmonic oscillator,
\begin{equation}
a_{\omega}^{\pm}=\frac{1}{\sqrt{2}}\left(\mp \frac{\text{d}}{\text{d} x}+\omega x\right),\label{fac1}
\end{equation}
it can be shown that (see also \cite{Shi00})
\begin{equation}\label{2wf}
H = a_{\omega}^{+}a_{\omega}^{-}+\frac{\omega}{2} = a_{\omega}^{-}a_{\omega}^{+}-\frac{\omega}{2}.
\end{equation}
Since for $\omega=1$ the operators $a_{1}^{+}$ and $a_{1}^{-}$ are mutually hermitian conjugate, $(a_{1}^{-})^{\dag}=a_{1}^{+}$, then for the harmonic oscillator the two factorizations of Eq.~\eqref{2wf} are essentially different. On the other hand, for $\omega=i$ the operators $a_{i}^{\pm}$ are antihermitian,  $(a_{i}^{\pm})^{\dag}=-a_{i}^{\pm}$, which implies that for the inverted oscillator the two factorizations in Eq.~\eqref{2wf} are indeed the same (since $H$ is hermitian).

It is clear now that the set of operators $\{H,a_{\omega}^{+},a_{\omega}^{-}\}$ satisfies the following algebra
\begin{align}
\left[H,a_{\omega}^{\pm}\right]& =\pm\omega a_{\omega}^{\pm},\\
\left[a_{\omega}^{-},a_{\omega}^{+}\right]&=\omega,
\end{align}
which leads immediately to
\begin{equation}
Ha_{\omega}^{\pm}\psi(x)=(E\pm \omega)a_{\omega}^{\pm}\psi(x),
\end{equation}
where $\psi(x)$ satisfies Eq.~\eqref{sseq}, and applying $n$ times $a_{\omega}^{\pm}$ we get
\begin{equation}
H(a_{\omega}^{\pm})^n\psi(x)=(E\pm n\omega)(a_{\omega}^{\pm})^n\psi(x).\label{fac6}
\end{equation}
Note that for both cases, $\omega=1$ and $\omega=i$, the general solution of Eq.~\eqref{sseq} is given by Eq.~\eqref{roy1} in the extended domain $E\in\mathbb{C}$; however, these solutions are not always physically admissible. Next, we will examine in more detail each of the two algebras.

\subsection{Harmonic oscillator algebra}

For $\omega=1$, Eqs.~(\ref{fac1}--\ref{fac6}) simplify to the Heisenberg-Weyl algebra of the harmonic oscillator potential \cite{per86}. The general solution of the SSE is given by Eq.~\eqref{roy1} for $\omega=1$, from which we can obtain its {\it bound states} or its {\it pure point spectrum}.


An alternative way is to take the first factorization of Eq.~\eqref{2wf}, $H=a_1^{+}a_1^{-}+1/2$, and look for the {\it extremal state} $\psi_0(x)$ which is annihilated by $a_1^{-}$,
\begin{equation}
a_1^{-}\psi_0(x)=0 \quad \Rightarrow \quad \psi_0(x)=\pi^{-1/4}\exp(-x^2/2).\label{base}
\end{equation}
In addition, $\psi_0(x)$ is an eigenfunction of $H$ with eigenvalue $E_0=1/2$, and if we apply iteratively the creation operator we will get the remaining bound states
\begin{equation}\label{excitados}
\psi_n(x)=\frac{(a_1^{+})^n}{(n!)^{1/2}}\psi_0(x)=\frac{1}{2^{n/2}\pi^{1/4}(n!)^{1/2}}\exp(-x^2/2)H_n(x),
\end{equation}
associated with the eigenvalues $E_n=n+1/2,\ n=0,1,\dots$, where $H_n(x)$ are the Hermite polynomials. This kind of algebra is known as {\it spectrum generating algebra}. Similarly, we can obtain another extremal state $\phi_0(x)$ from the second factorization, $H=a_1^{-}a_1^{+}-1/2$, as
\begin{equation}
a_1^{+}\phi_0(x)=0 \quad \Rightarrow\quad \phi_0(x)=\pi^{-1/4}\exp(x^2/2),
\end{equation}
which is a solution of the SSE associated with $E =-1/2 \equiv e_0$. We have added the constant factor $\pi^{-1/4}$ by symmetry with Eq.~\eqref{base}, even though the wavefunction $\phi_0(x)$ is not normalizable and therefore it is not a bound state. We have stressed this fact by choosing a different notation $\phi_0$ for this wavefunction.

Note that the extremal state $\phi_0(x)$ can also be obtained by acting $a_1^-$ on the {\it irregular wavefunction} $\varphi_0(x)$, defined as the second linearly independent solution of the SSE for $E_0=1/2$. Although this wavefunction is not normalizable, it has some physical applications \cite{Leo10}.

\begin{figure}
\begin{center}
\includegraphics[scale=0.38]{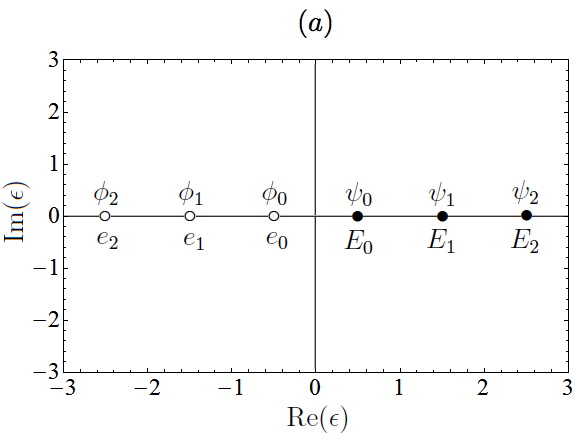}
\includegraphics[scale=0.38]{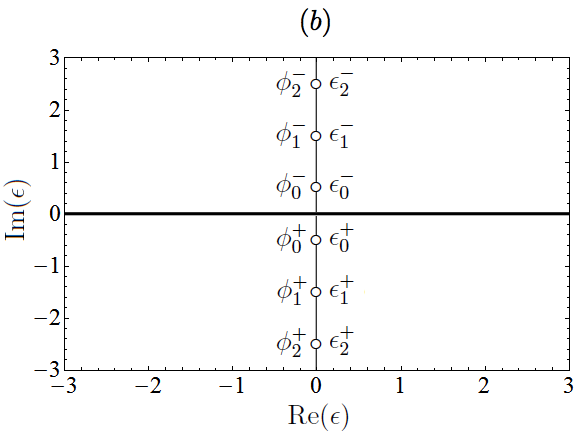}
\end{center}
\vspace{-5mm}
\caption{The complex plane for which a general solution $\forall\, E\in\mathbb{C}$ can be found for both the harmonic and the inverted oscillators. In (a) we show the bound states (black dots) and the polynomial solutions (white dots) for the harmonic oscillator. In (b) we show the scattering states on the real axis (black line) and the polynomial solutions (white dots) for the inverted oscillator.} \label{complexplane}
\end{figure}

Now, we will apply iteratively the annihilation operator in order to obtain a new ladder of wavefunctions expressed in terms of Hermite polynomials, although with a different argument as compared with the standard case. Hence, the ladder of nonphysical wavefunctions is given by
\begin{equation}\label{enesimosnofisicos}
\phi_n(x)=\frac{(a_1^{-})^n}{(n!)^{1/2}}\phi_0(x)=\frac{i^{-n}}{2^{n/2}\pi^{1/4}(n!)^{1/2}}\exp(x^2/2)H_n(ix),
\end{equation}
associated with the discrete nonphysical energies $e_n=-n-1/2,\ n=0,1,\dots$, which nevertheless provides an additional set of {\it polynomial solutions} for the oscillator potential. The factor $i^{-n}$ appears naturally when we factorize the Hermite polynomial with the correct argument. The position of the eigenvalues for the bound states \eqref{excitados} and the nonphysical energies associated with the polynomial solutions \eqref{enesimosnofisicos} on the complex $E$-plane can be seen in Fig.~\ref{complexplane}(a).

\subsection{Inverted oscillator}

For $\omega=i$ Eqs.~(\ref{fac1}--\ref{fac6}) define a zeroth order complex deformation of the Heisenberg-Weyl algebra. The general solution of the corresponding SSE is given by Eq.~\eqref{roy1} with $\omega=i$ in the full complex $E$-plane.

In this case there is no solution which fulfills the bound state condition, as it is expected because the inverted oscillator potential has neither local nor global wells which would support such bound states. Nevertheless, it is expected to have {\it dispersive states} that, under certain conditions, can be given a physical interpretation. Note that the most familiar system with these characteristic states is the free particle.

As it was shown in Section 2, the natural dispersive states of the inverted oscillator are those given by  
Eqs.~(\ref{uplus}--\ref{uminus}) and (\ref{ul}--\ref{ur}) for any $E\in\mathbb{R}$. However, we can ask ourselves further: Are there solutions related with the Hermite polynomials for the inverted oscillator? In order to answer, let us find now the extremal states associated with the factorizations of Eq.~\eqref{2wf} for $\omega=i$.

Let us consider in the first place
\begin{equation}
a_i^{-}\phi_0^{-}(x)=0 \quad \Rightarrow \quad \phi_0^{-}(x)=\pi^{-1/4}\exp(-ix^2/2),
\end{equation}
for $\epsilon_0^{-}=i/2$. Furthermore, if we apply iteratively the creation operator $a_i^{+}$ we will get the following set of polynomial solutions of the SSE
\begin{equation}\label{enesimosnofisicosinvertedmenos}
\phi_n^{-}(x)=\frac{(a_i^{+})^n}{(n!)^{1/2}}
\phi_0^{-}(x)=\frac{i^{n/2}}{2^{n/2}\pi^{1/4}(n!)^{1/2}}\exp(-ix^2/2)H_n(i^{1/2}x),
\end{equation}
associated with $\epsilon_n^{-}=i(n+1/2),\ n=0,1,\dots$

Similarly, the use of the second factorization leads to
\begin{equation}
a_i^{+}\phi_0^{+}(x)=0 \quad \Rightarrow \quad \phi_0^{+}(x)=\pi^{-1/4}\exp(ix^2/2),
\end{equation}
associated with $\epsilon_0^{+}=-i/2$. In addition, if we apply several times the annihilation operator $a_i^{-}$ we will get another set of polynomial solutions
\begin{equation}\label{enesimosnofisicosinvertedmas}
\phi_n^{+}(x)=\frac{(a_i^{-})^n}{(n!)^{1/2}}
\phi_0^{+}(x)=\frac{i^{-n/2}}{2^{n/2}\pi^{1/4}(n!)^{1/2}}\exp(ix^2/2)H_n(i^{3/2}x),
\end{equation}
associated with $\epsilon_n^{+}=-i(n+1/2),\ n=0,1,\dots$.

Let us stress once again that $\phi_n^{-}(x)$ and $\phi_n^{+}(x)$ are solutions of an SSE for complex $E$-values and thus they do not have any physical  interpretation at all, neither as bound nor as dispersive states. Nevertheless, it turns out that they are related with polynomial solutions of the SSE. A diagram marking the positions in which this kind of solutions appear for the inverted oscillator on the complex $E$-plane is shown in Fig.~\ref{complexplane}(b).

Note that there is a direct extension of this factorization method, which is closely related to SUSY QM. In this generalization, instead of factorizing the Hamiltonian $H$ in two different ways in terms of a pair of first-order operators, as in Eq.~\eqref{2wf}, one looks for the most general first-order operators producing just a single factorization \cite{Mie84}. Thus, when the ordering of the operators factorizing $H$ is reversed in general one arrives to a different Hamiltonian. This fact has been used to generate new exactly-solvable Hamiltonians departing from a given initial one \cite{Mie84,fe84,Suk85}.

Next, we are going to apply the first- and second-order SUSY QM to the inverted oscillator. This supplies us with the building bricks for implementing the higher-order transformations, since it is known nowadays that any non-singular transformation of order higher than two can always be factorized in terms of non-singular first- and second-order SUSY transformations. Note that, although this fact was conjectured for the first time in \cite{aicd95,bs97}, however it was proved in a rigorous way just recently \cite{as07,so08}. Then, the higher-order SUSY partners of the inverted oscillator can be obtained through iterations of the non-singular transformations which will be discussed here.

\section{First-order SUSY QM}

In the first-order SUSY QM one starts from a given solvable Hamiltonian $H_0$ and tries to
factorize it directly in terms of two first-order operators. An alternative method is to depart from
an equivalent intertwining relationship and to solve the resulting system of equations. We will use the last
procedure starting from the inverted oscillator Hamiltonian
\begin{equation}
H_0=  -\frac{1}{2}\frac{\text{d}^2}{\text{d} x^2} + V_0(x)= -\frac{1}{2}\frac{\text{d}^2}{\text{d} x^2}-\frac{1}{2}x^2,
\end{equation}
and looking for a first-order differential operator $A_1^{+}$ which intertwines $H_0$ with a new
Hamiltonian $H_1$ in the way \cite{Mie84,fe84,Suk85}
\begin{equation}
H_1A_1^{+}=A_1^{+}H_0,
\end{equation}
where
\begin{align}
A_1^{+} &=\frac{1}{\sqrt{2}}\left[-\frac{\text{d}}{\text{d} x}+\alpha_1(x)\right],\\
H_1 &= - \frac{1}{2}\frac{\text{d}^2}{\text{d} x^2}+V_1(x).
\end{align}
In this formalism $\alpha_1(x)$ is called the {\it superpotential} and $V_1(x)$ denotes the new potential; until now, both functions are still to be determined.

Using the standard results of the first-order SUSY QM, it turns out that $\alpha_1(x)$ must
satisfy the Riccati equation
\begin{equation}\label{1riccati}
\alpha_1'+\alpha_1^2=2(V_0-\epsilon).
\end{equation}
Through the substitution $\alpha_1=u'/u$, Eq.~\eqref{1riccati} is transformed into
\begin{equation}
- \frac{1}{2}u''(x)+V_0(x)u(x)=\epsilon u(x).
\end{equation}
This is the SSE for the initial potential with $E=\epsilon$ whose general solution is given by Eq.~\eqref{roy2}. Thus, it is straightforward to obtain the formula for the new potential as
\begin{equation}
V_1(x) = V_0(x) - \alpha_1'(x)= V_0(x)-\left[\frac{u'(x)}{u(x)}\right]',
\end{equation}
which is real for any $\epsilon, C, D\in{\mathbb R}$. It is clear now that if the {\it transformation
function} $u(x)$ has zeros, then the new potential $V_1(x)$ will have singularities at those points. Since the seed solution $u(x)$ given in Eq.~\eqref{roy2}, associated with an arbitrary {\it factorization energy} $\epsilon\in{\mathbb R}$, always has zeros because it has oscillatory terms that cannot get cancelled (see Fig.~\ref{figure1}), it follows that it is impossible to perform real non-singular first-order SUSY transformations for the inverted oscillator. Note that the singular transformations are excluded because they change, in general, the domain of the initial potential and, consequently, the initial spectral problem (compare, e.g., \cite{mnn98}).

\section{Second-order SUSY QM}

The second-order SUSY QM typically starts from the following intertwining relationship \cite{Fer10,aicd95}
\begin{equation}
H_2 B^{+} = B^{+} H_0,\label{2S_Entrelaza}
\end{equation}
where now
\begin{align}
H_2 &= - \frac12\frac{\text{d}^2}{\text{d} x^2} + V_2(x),\label{2S_Ht}\\
B^{+} &= \frac{1}{2}\left[\frac{\text{d}^2}{\text{d} x^2} - g(x)\frac{\text{d}}{\text{d} x}+h(x)\right].\label{2S_Bmas}
\end{align}
The aim is to determine $V_2(x)$, $g(x)$ and $h(x)$ supposing that some solutions of the SSE for the potential $V_0(x)$ are known. First of all, the unknown functions $V_2(x)$ and $h(x)$ are expressed in terms of $V_0(x)$ and $g(x)$ in the way
\begin{align}
V_2(x) & = V_0 - g',\label{eqv} \\
h(x) & = \frac{g'}{2} +\frac{g^2}{2} - 2V_0 + d. \label{hcom}
\end{align}
Next, it turns out that $g(x)$ must fulfill the following non-linear second-order differential equation:
\begin{equation}
\frac{gg''}{2}-\frac{g'^2}{4}+g^2\left(g'+\frac{g^2}{4}-2V_0+d \right)+c=0, \label{eqg}
\end{equation}
with $c,d\in\mathbb{R}$ being two integration constants. In order to solve Eq.~\eqref{eqg}, let us
use the ansatz \cite{FGN98}
\begin{equation}
g' = -g^2+2\gamma g + 2 \xi, \label{ans}
\end{equation}
where $\gamma(x)$ and $\xi(x)$ are functions to be determined. Substituting this ansatz in
Eq.~\eqref{eqg}, it is obtained that $\xi^2=c$ and the following Riccati equation for $\gamma(x)$:
\begin{equation}
\gamma' + \gamma^2 = 2(V_0 - \epsilon),\label{2S_Ricatti}
\end{equation}
with $\epsilon = (d+\xi)/2$. Therefore, to determine the function $\gamma$ one just has to solve
Eq.~\eqref{2S_Ricatti}, which can be linearized by using $\gamma=u'/u$, leading to an SSE:
\begin{equation}
-\frac12u'' + V_0 u = \epsilon u. \label{sch}
\end{equation}
The employed seed solutions $u$ are called once again {\it transformation functions} and the associated $\epsilon$ {\it factorization energies}. The corresponding transformations, leading to final real potentials $V_2(x)$, depend as well on the sign of $c$. Thus, they can be classified in three different cases: real, confluent, and complex (see Table~\ref{table1}).
\begin{table}
\begin{center}
\begin{tabular}{ll}
\hline
Value of $c$&Type of transformation\\
\hline
$c>0$&Real case\\
$c=0$&Confluent case\\
$c<0$&Complex case\\
\hline
\end{tabular}
\end{center}
\vspace{-3mm}
\caption{The three types of second-order SUSY transformations.} \label{table1}
\end{table}

\subsection{The real case for $c>0$}
In this case it turns out that $\xi_1 = \sqrt{c}>0, \ \xi_2 = - \sqrt{c}$, $\epsilon_i = (d + \xi_i)/2 \in {\mathbb R}, \ i=1,2,$ $\epsilon_1\neq\epsilon_2$. The corresponding known solutions of the Riccati equation \eqref{2S_Ricatti} are denoted as $\gamma_1(x)$,
$\gamma_2(x)$, and each one of them leads to (see ansatz \eqref{ans})
\begin{align}
g'(x) = - g^2(x)+2\gamma_1(x)g(x)+2(\epsilon_1-\epsilon_2),\\
g'(x) = - g^2(x)+2\gamma_2(x)g(x)+2(\epsilon_2-\epsilon_1).
\end{align}
By subtracting them we arrive at
\begin{equation}
g(x)= -\frac{2(\epsilon_1 - \epsilon_2)}{\gamma_1 - \gamma_2} = \{\log[W(u_1,u_2)]\}',
\end{equation}
where $u_1, \ u_2$ are two solutions of the initial stationary Schr\"odinger equation associated with $\epsilon_1, \ \epsilon_2$ and
$W(u_1,u_2) = u_1u_2' - u_1'u_2$ is their Wronskian. The new potential $V_2(x)$ is given by
\begin{equation}
V_2(x)=V_0(x)-\{\log[W(u_1,u_2)]\}'',\label{v2}
\end{equation}
i.e., to obtain a non-singular potential $V_2$ a $W(u_1,u_2)$ without zeros is required (recall that in the first-order case it was directly the transformation function $u$ the one that should not have nodes). A $W(u_1,u_2)$ without zeros could be achieved if $u_1$ and $u_2$ would have alternate nodes. For the inverted oscillator this requirement is true in some finite interval of the $x$-domain but not in the full real line. Therefore, we cannot produce real non-singular second-order SUSY transformations for the inverted oscillator. On the other hand, note that the zeros of $u_1$, $u_2$ are closer to alternate in all ${\mathbb R}$ as $\epsilon_1$ and $\epsilon_2$ become closer. This fact hints us to use the confluent second-order SUSY QM, a well worked algorithm where the two factorization energies converge to a single one \cite{bff12,mnr00,FSH03}.

\subsection{The confluent case for $c=0$}
\label{confluent}
In this case $\epsilon_1=\epsilon_2:=\epsilon\in\mathbb{R}$, and thus the ansatz in Eq.~\eqref{ans} leads
to a Bernoulli equation for $g$:
\begin{equation}
g' - 2\gamma g = - g^2.
\end{equation}
Let us suppose that $\gamma = u'/u$ is known. Thus, the general solution for $g$ becomes
\begin{equation}
g(x)=\{\log[w(x)]\}',
\end{equation}
with
\begin{equation}
w(x)=w_0+\int_{x_0}^{x}u^2(z)dz,\label{omega}
\end{equation}
where $w_0$ is a constant and $x_0$ is a fixed point inside the domain of $V_0$. In order to obtain a non-singular $V_2(x)$ we
need to employ a $w(x)$ without zeros. Since $w(x)$ is a monotonically non-decreasing function, one can use a transformation function \cite{FSH03} such that
\begin{equation}
\lim_{x\rightarrow \infty} u(x)=0, \quad \nu_+ = \int_{x_0}^{\infty}|u(z)|^2dz<\infty, \label{cond1}
\end{equation}
or
\begin{equation}
\lim_{x\rightarrow -\infty} u(x)=0, \quad \nu_- = \int_{-\infty}^{x_0}|u(z)|^2dz<\infty. \label{cond2}
\end{equation}
This means that $u(x)$ should have a null asymptotic behaviour in one boundary and also be square-integrable over the corresponding semi-bounded interval.

Now, from the asymptotic behaviour of $\psi_e$ and $\psi_o$ given in Eqs.~(\ref{limue}--\ref{limuo}) we can see that the leading terms fall off as $\sim |x|^{-1/2}$ (remarkably they do not depend on $\epsilon\in\mathbb{R}$); therefore they are in the frontier (but outside) of $\mathcal{L}^2(\mathbb{R})$, as any improper Dirac base. This means that they fulfill the first conditions of Eqs.~\eqref{cond1} and \eqref{cond2} but not the second ones, i.e., there is no linear combination of solutions which can be made square-integrable over a semi-bounded interval. This fact implies that the $w(x)$ of Eq.\eqref{omega} will always have one zero on the real axis, and consequently the confluent second-order SUSY transformation will always be singular. Thus one can conclude that, although at first sight it seemed possible to perform the non-singular SUSY transformation through the confluent algorithm, however the transformation functions $u(x)$ obtained from Eq.~(\ref{roy2}) by making $E=\epsilon$ do not fulfill neither Eqs.~\eqref{cond1} nor Eqs.~\eqref{cond2}.

\subsection{The complex case for $c<0$}
For $c<0$ it turns out that  $\epsilon_1,\epsilon_2\in\mathbb{C}$ and $\epsilon_1=\overline{\epsilon}_2:=\epsilon$. In this case we are only interested in those transformations producing a real potential $V_2(x)$, which implies that $\gamma_1(x)=\overline{\gamma}_2(x) := \gamma(x)$. Following a procedure similar to the real case, we obtain \cite{FMR03,RM03}
\begin{equation}
g(x)=\{\log [w(x)]\}'=\frac{w'}{w}=\frac{u\overline{u}}{w}, \label{gcom}
\end{equation}
where now
\begin{equation}
w(x)=\frac{W(u,\overline{u})}{2(\epsilon-\overline{\epsilon})}.\label{funw}
\end{equation}
Furthermore, it is easy to show that $w(x)=\overline{w}(x)$, i.e., $w(x)\in \mathbb{R}$.

Once again, $w(x)$ must not have zeros in ${\mathbb R}$ to avoid singularities in $V_2(x)$. Since $w'(x)=|u(x)|^{2}$, then $w(x)$ is a monotonically non-decreasing function. Thus, to assure that $w(x)\neq 0\ \forall \ x\in\mathbb{R}$ it is sufficient to fulfill either the conditions of Eqs.~\eqref{cond1} or those of Eqs.~\eqref{cond2}, although both cannot be accomplished since the transformation function $u(x)$ used to implement the SUSY algorithm cannot be physical, i.e., it is not square-integrable. The complex second-order transformations such that $u(x)$ fulfills one of these conditions can produce real potentials which are always strictly isospectral to the original one.

As it was discussed at the end of Section 5.2, for real factorization energies $\epsilon\in\mathbb{R}$ we cannot find a transformation function $u(x)$ that fulfills either Eqs.~\eqref{cond1} or Eqs.~\eqref{cond2}. On the other hand, from the asymptotic behaviour for $\psi_e(x)$ and $\psi_o(x)$ given in Eqs.~(\ref{limue},\ref{limuo}), we can see that both functions contain two terms, one going as $x^{-1/2-i\epsilon}$ and other as $x^{-1/2+i\epsilon}$. Then, for $\mbox{Im}(\epsilon)\neq 0$ we will have one term falling faster and other slower than $x^{-1/2}$. Moreover, if a linear combination of the two solutions which cancels the slower terms can be found, thus a solution that fulfills either the conditions of Eqs.~\eqref{cond1} or of Eqs.~\eqref{cond2} is obtained.

\begin{figure}
\begin{center}
\includegraphics[scale=0.5]{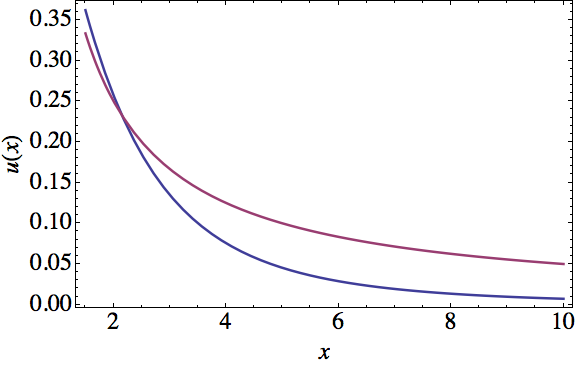}
\end{center}
\vspace{-5mm}
\caption{Comparison between $|u_P(x,\epsilon)|^2$ for $\epsilon=5+i$ (blue line) and $\propto x^{-1}$ (magenta line). We can see here that the conditions of Eq.~\eqref{cond1} are fulfilled.} \label{figu}
\end{figure}

It is straightforward to obtain now the linear combinations that fulfill the condition of Eqs.~(\ref{cond1}), the resulting functions being given by
\begin{align}
u_P(x,\epsilon)&=\psi_e(x) - \frac{2\, \text{e}^{-i\pi /4} \Gamma(3/4-i \epsilon/2)}{\Gamma(1/4 - i \epsilon /2)}\psi_o(x),\label{up}\\
u_N(x,\epsilon)&=\psi_e(x) - \frac{2\, \text{e}^{i\pi /4} \, \Gamma(3/4 + i \epsilon/2)}{\Gamma(1/4 + i \epsilon /2)}\psi_o(x),
\end{align}
where the labels $P$ and $N$ refer to `positive' and `negative', according to the sign of $\mbox{Im}(\epsilon)$ which gives the right behaviour, and we have expressed explicitly the dependence of those functions from the complex factorization energy $\epsilon$. In Fig.~\ref{figu} it is shown a graph of $|u_P(x,\epsilon)|^2$ as compared with $\propto x^{-1}$. In this case the conditions of Eq.~\eqref{cond1} are fulfilled and thus we can perform the non-singular complex SUSY transformation. Note that the transformation function $u_P(x,\epsilon)$ of Eq.~\eqref{up}, which is square-integrable in $\mathbb{R}^{+}$ for $\mbox{Im}(\epsilon)>0$, coincides with the solution found in Section 20.10 of \cite{Tit58}.

\begin{figure}
\begin{center}
\includegraphics[scale=0.5]{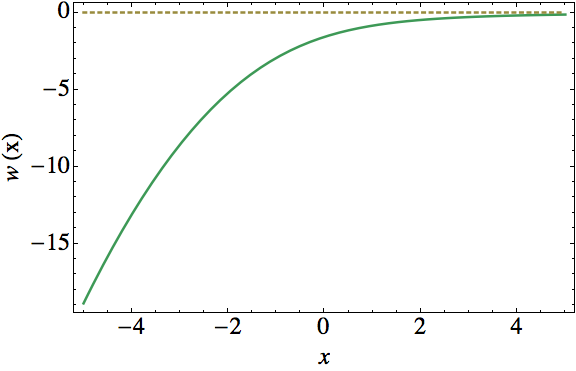}
\end{center}
\vspace{-5mm}
\caption{The function $w(x)$ for the complex `energy' $\epsilon=5+i$ (green line). The zero of this function lies in $+\infty$.} \label{figw}
\end{figure}

Next, we need to construct the function $w(x)$ defined by Eq.~\eqref{funw}, which is a real-defined function. In order to obtain a real non-singular SUSY partner potential \cite{FMR03}, this $w(x)$ should be built up by using only either $u_{P}(x,\epsilon)$ or $u_{N}(x,\epsilon)$ as transformation functions, which makes that the non-decreasing monotonic function $w(x)$ vanishes either at $x\rightarrow \infty$ if Eqs.~(\ref{cond1}) are obeyed or at $x\rightarrow -\infty$ if Eqs.~(\ref{cond2}) are satisfied. In Fig.~\ref{figw} we show the function $w(x)$ built from $u_{P}(x,\epsilon)$ for a specific factorization energy.

\begin{figure}
\begin{center}
\includegraphics[scale=0.6]{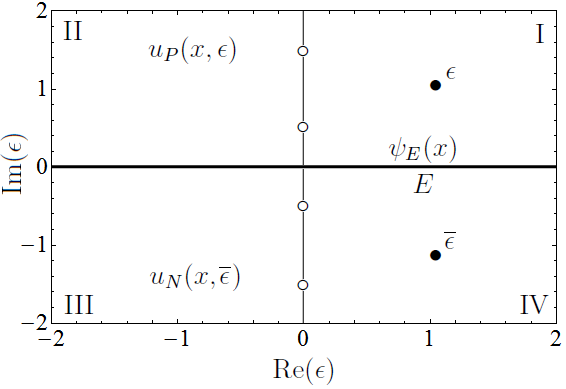}
\end{center}
\vspace{-5mm}
\caption{Domain of the transformation functions $u_P(x,\epsilon)$ and $u_N(x,\overline{\epsilon})$ in the complex plane $\epsilon$. The real line ($\mbox{Im}(\epsilon)=0$) and the points $\epsilon = \pm i(m + 1/2), \ m=0,1,2,\dots$ are excluded. For quadrants I and II we should use $u_P(x,\epsilon)$ as transformation function, and $u_N(x,\epsilon)$ for quadrants III and IV.} \label{figure4}
\end{figure}

This analysis shows that we can implement successfully the non-singular complex SUSY transformations through any of the two seed solutions $u_P(x,\epsilon)$ or $u_N(x,\epsilon)$. Nevertheless, we should remember that each of these functions works only for half of the complex plane $\epsilon$, i.e., the imaginary part of $\epsilon$ should be positive in order to use $u_P(x,\epsilon)$ or negative for $u_N(x,\epsilon)$. However, by noticing that
\begin{equation}
{\overline u}_P(x,\epsilon) = u_N(x,{\overline \epsilon}),
\end{equation}
and looking at the algorithm more carefully it is  seen that for $\mbox{Im}(\epsilon)> 0$ both the complex factorization energy $\epsilon$ (with its corresponding transformation function $u_P(x,\epsilon)$) and its complex conjugate ${\overline \epsilon}$ (with ${\overline u}_P(x,\epsilon)= u_N(x,{\overline \epsilon})$) are really used by the transformation. This means that all possible non-singular transformations are already covered by using either $u_P(x,\epsilon)$ or $u_N(x,\epsilon)$ with $\epsilon$ lying on the right domain, as it is shown in Fig.~\ref{figure4}.

Note that the real line of the complex plane $\epsilon$ is excluded from the domain of non-singular SUSY transformations since our previous analysis is valid just for $\mbox{Im}(\epsilon)\neq 0$. Furthermore, the points $\epsilon = \pm i(m + 1/2), \ m=0,1,2,\dots$ are also excluded because at those points the solutions reduce to the polynomials of Section 3.2 and the asymptotic behaviour either of $\psi_e$ or $\psi_o$, given by Eqs.(\ref{limue},\ref{limuo}), is no longer valid.

The analytic expression for the new potential $V_2$ can be found by substituting Eqs. (\ref{gcom},\ref{funw}) into Eq.~\eqref{eqv} in order to obtain
\begin{equation}
V_2(x)=-\frac{x^2}{2}-\left(\frac{u\overline{u}' + u'\overline{u}}{w}-\frac{(u\overline{u})^2}{w^2}\right).
\end{equation}

\begin{figure}
\begin{center}
\includegraphics[scale=0.6]{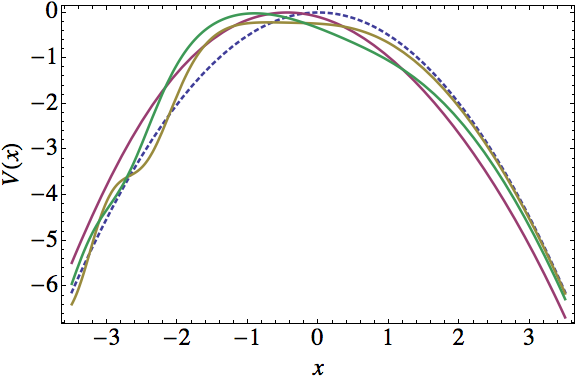}
\end{center}
\vspace{-5mm}
\caption{Inverted oscillator potential (blue dashed line) and its second-order SUSY partners generated by using $u_P(x,\epsilon)$ as transformation function for different complex values of $\epsilon$, $\{ 10^{-5}+5i,(1+i)/5,10^{-2}+i \}$, which correspond to the green, magenta, yellow lines respectively.} \label{figure5}
\end{figure}

In Fig.~\ref{figure5} we show several supersymmetric partners of the inverted oscillator potential, built through this complex algorithm. We want to remark that, in these cases, there are no new bound states at all being created by the transformations.

\section{Algebra of the new Hamiltonians}

In the last Subsection we have finally obtained new potentials through SUSY QM departing from the inverted oscillator. In fact, they constitute a two-parametric family of potentials, whose parameters are $\mbox{Re}(\epsilon)$, $\mbox{Im}(\epsilon)$. We have also shown that $\epsilon$ and $\overline{\epsilon}$ induce the same transformation and, consequently, the same potential $V_2(x)$.

Remember that $H_2$ is isospectral to $H_0$ for the only second-order SUSY transformation that works, namely, for the complex case. Note that $B^{-}\equiv (B^{+})^{\dag}$, then it turns out that
\begin{align}
B^{+}B^{-}=(H_2-\epsilon)(H_2-\overline{\epsilon}),\\
B^{-}B^{+}=(H_0-\epsilon)(H_0-\overline{\epsilon}).
\end{align}

Let us define now a pair of new operators $L_i^{\pm}$ as
\begin{equation}
L_i^{\pm}\equiv B^{+}a_i^{\pm}B^{-},
\end{equation}
which act onto the eigenfunctions of the new Hamiltonian $H_2$ as ladder operators because they satisfy the following algebra

\begin{figure}
\begin{center}
\includegraphics[scale=0.4]{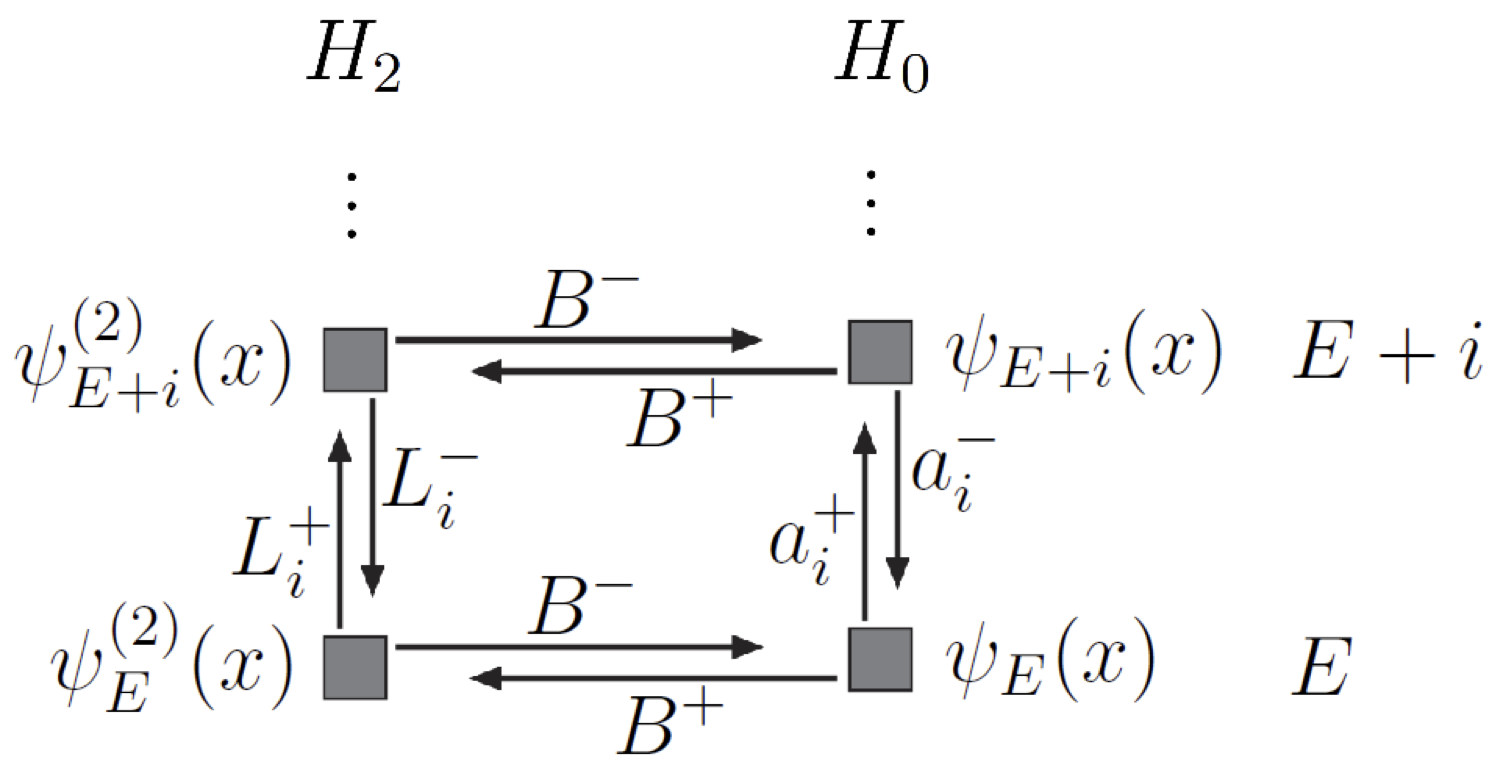}
\end{center}
\vspace{-5mm}
\caption{Diagram of the complex second-order SUSY transformation. $H_0$ has ladder operators $a_i^{\pm}$, and $H_2$ has also ladder operators denoted by $L_i^{\pm}$, which are built from $B^{\pm}$ and $a_i^{\pm}$.} \label{figsusy2}
\end{figure}

\begin{align}
[H_2,L_i^{\pm}]&=\pm i L_i^{\pm},\\
[L_i^{-},L_i^{+}]&= P_5(H_2+i)-P_5(H_2)=Q_4(H_2),
\end{align}
where $Q_4,P_5$ are complex polynomials of orders $4$ and $5$ respectively. They are given by the analogue of the number operator for $H_2$:
\begin{align}
\hskip-1cm L_i^{+}L_i^{-} = (H_2-\epsilon)(H_2-\overline{\epsilon})(H_2-\epsilon-i)(H_2-\overline{\epsilon}-i)(H_2-i/2) \equiv P_5(H_2) .
\end{align}
Note that the ladder operators $L_i^{\pm}$ of $H_2$ are also antihermitian, $(L_i^{\pm})^\dagger = - L_i^{\pm}$, as it happens for the inverted oscillator.

\begin{figure}
\begin{center}
\includegraphics[scale=0.612]{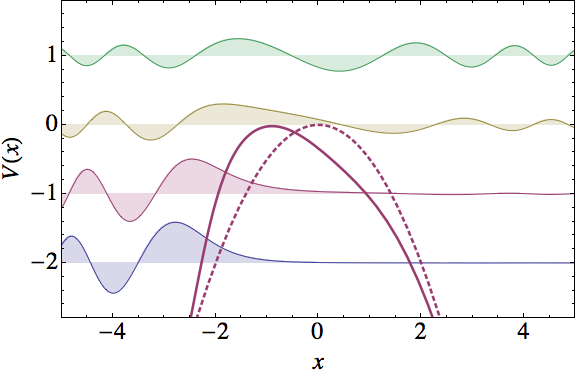}
\end{center}
\vspace{-5mm}
\caption{SUSY partner potential (solid curve) of the inverted oscillator (dashed line) and four of its eigenfunctions associated with the energies $E\in\{-2,-1,0,1\}$. The factorization energy involved in this transformation is $\epsilon = 10^{-5}+5i$.} \label{fignew}
\end{figure}

Next we can obtain the analytic expression for the eigenfunctions of the new Hamiltonian $H_2$. Using Eq.~\eqref{2S_Entrelaza} we will get \cite{ff05}
\begin{align}
B^{+}\psi_E & =\sqrt{(E-\epsilon)(E-\overline{\epsilon})}\psi_E^{(2)}, \label{bmastimesb}\\
B^{-}\psi_E^{(2)} & =\sqrt{(E-\epsilon)(E-\overline{\epsilon})}\psi_E,
\end{align}
where $\psi_E(x)$ denotes an eigenfunction of $H_0$ associated with an arbitrary real energy $E$ (it can be the $\psi_E^\pm(x)$ of Eqs.~(\ref{uplus}-\ref{uminus}), the $\psi_L(x)$ of Eq.~(\ref{ul}), the $\psi_R(x)$ of Eq.~(\ref{ur}) or a linear combination of both) and $\psi^{(2)}_E(x)$ denotes the corresponding eigenfunction of $H_2$ (see diagram in Fig.~\ref{figsusy2}). Indeed, by substituting in the definition of $B^{+}$ (see Eq.~(\ref{2S_Bmas})) the expressions for $g(x)$ and $h(x)$ of Eqs.~(\ref{hcom},\ref{gcom}) and after several simplifications we get
\begin{equation}
\psi_E^{(2)}(x) \propto \frac{w'(x)}{w(x)} \left[ - \psi_E'(x) + \frac{u'(x)}{u(x)}\psi_E(x) \right] + 2(\epsilon - E) \psi_E(x).
\end{equation}
We should recall that $u(x)$ is a complex transformation function, $\epsilon\in\mathbb{C}$ is the corresponding factorization energy and $E\in\mathbb{R}$ is the energy associated with the initial and transformed eigenfunctions $\psi_E(x),\psi_E^{(2)}(x)$. Note that the asymptotic behaviour for $\psi_E^{(2)}(x)$ is the same as for the initial eigenfunction. In Fig.~\ref{fignew} a new potential $V_2(x)$ and four of its associated eigenfunctions are shown.

\section{Conclusions}

In this work, the supersymmetric quantum mechanics has been successfully employed to generate new exactly-solvable potentials departing from the inverted oscillator. We have shown that the first-order as well as the real and confluent second-order SUSY transformations always produce singular potentials. On the other hand, through the complex second-order SUSY QM it is possible to generate new real non-singular exactly-solvable potentials.

It has been shown that this transformation can be achieved through two specific complex seed solutions which are the only ones fulfilling the appropriate conditions in order to produce new non-singular real potentials isospectral to the inverted oscillator. Furthermore, we have obtained a simple analytic expression for the eigenfunctions associated with the new Hamiltonian $H_2$. Let us note that the potentials generated in this paper can be used as models in every physical situation where the inverted oscillator has been employed before (see \cite{Bar86,YKC06,GP91,Shi00}). This is because the new Hamiltonians are isospectral to the original one and the form of their potentials is quite similar. These facts, in particular, could be important to foresee alternative models for describing the small imperfections appearing when a real Penning trap is built up \cite{bg86,cf11,cfv11}, specially if they do not change the spectrum of the corresponding ideal arrangement.

Furthermore, we have studied the general algebraic structure of the original system with arbitrary $\omega$, which is reduced to the harmonic oscillator case for $\omega=1$. We have analyzed in detail as well the case with $\omega=i$, which is related to the inverted oscillator and turns out to have a Heisenberg-Weyl deformed complex algebraic structure. For the new Hamiltonians obtained through the complex second-order SUSY QM applied to the inverted oscillator we have examined as well the related algebra in some detail.

In the future we would like to analyze further the algebras associated with a more general system with an arbitrary $\omega$, different from $1$ and $i$, since the symmetry in the wavefunctions $\psi_n,\phi_n,\phi_n^{-},\phi_n^{+}$ of Eqs.~(\ref{excitados},\ref{enesimosnofisicos},\ref{enesimosnofisicosinvertedmenos},\ref{enesimosnofisicosinvertedmas}) suggests the existence of a common structure that, perhaps, would allow us to understand deeper this family of oscillators.

\section*{Acknowledgement}
The authors acknowledge the financial support of Conacyt project 152574, as well as the useful comments and suggestions of Professor Alexander Turbiner and the anonymous Referee of this paper. DB also acknowledges the Conacyt PhD scholarship 219665.

\newpage

\section*{Appendix}

In this Appendix we are going to derive the orthogonality and completeness relations
\begin{eqnarray}
&& (\psi_E^\sigma,\psi_{E'}^{\sigma'}) = \int_{-\infty}^{\infty} \overline{\psi}_E^{\, \sigma}(x)\psi_{E'}^{\sigma'}(x) dx 
=  \delta(E-E') \delta_{\sigma,\sigma'} , \label{a0} \\
&& \sum_{\sigma = \pm} \int_{-\infty}^{\infty} dE \psi_E^\sigma(x) \overline\psi_E^{\, \sigma}(x') =  \delta(x-x'), \label{a00}
\end{eqnarray}
for the eigenfunctions of the inverted oscillator Hamiltonian given in Eqs.~(\ref{uplus}--\ref{uminus}), where
\begin{eqnarray}
&& N_E = \frac{e^{i(1/2 - iE)\pi/4} \, 2^{iE/2-1} \, \Gamma(1/2-iE)}{\pi^{1/2}\, \Gamma(3/4 - iE/2)} , \nonumber
\end{eqnarray}
$\delta_{\sigma,\sigma'}$ is the Kronecker delta function in the indices $\sigma$ and $\sigma'$ and $\delta(y - y')$ is the Dirac delta function in the index $y$. We will point out just the main steps of the derivation given by Wolf (\cite{wo79}, Section 8.2).

First of all, departing from the standard Fourier transform, we introduce the bilateral Mellin transform $f_\sigma^{BM}(\lambda)$ of the function $f(x)$ and its inverse by means of
\begin{eqnarray}
&& f(x) = (2\pi)^{-1/2} \sum_{\sigma = \pm} \int_{-\infty}^\infty d\lambda \, f_\sigma^{BM}(\lambda) \, x_\sigma^{i\lambda - 1/2},  \label{a1} \\
&&  f_\sigma^{BM}(\lambda) = (2\pi)^{-1/2} \int_{-\infty}^\infty dx \, f(x) \, x_\sigma^{-i\lambda - 1/2}, \qquad \sigma = \pm, \label{a2}
\end{eqnarray}
where
\begin{eqnarray}
&& x_+ = 
\begin{cases}
x & {\rm for} \  x>0, \\
0 & {\rm for} \  x<0,
\end{cases} \label{a3} \\
&& x_- = \begin{cases}
0 & {\rm for} \  x>0, \\
-x & {\rm for} \  x<0.
\end{cases}  \label{a4}
\end{eqnarray}
By substituting now Eq.~(\ref{a1}) in Eq.~(\ref{a2}) the following orthogonality relation is obtained:
\begin{eqnarray}\label{a5}
&& (2\pi)^{-1} \int_{-\infty}^\infty dx \, x_\sigma^{-i\lambda - 1/2} \, x_{\sigma'}^{i\lambda' - 1/2} =  
\delta(\lambda - \lambda') \delta_{\sigma,\sigma'} .
\end{eqnarray}

On the other hand, if Eq.~(\ref{a2}) is substituted in Eq.~(\ref{a1}) one arrives at the following completeness relation:
\begin{eqnarray}\label{a6}
&& (2\pi)^{-1}\sum_{\sigma = \pm} \int_{-\infty}^{\infty} d\lambda \, x_\sigma^{i\lambda - 1/2} \, {x'}_\sigma^{-i\lambda - 1/2}
= \delta(x-x'),
\end{eqnarray}
meaning that the set of functions $\{(2\pi)^{-1/2} \, x_\sigma^{i\lambda - 1/2}, \ \sigma = \pm, \ -\infty<\lambda<\infty\}$ constitutes a generalized (Dirac) orthonormal basis for $\mathcal{L}^2(\mathbb{R})$.

Basically, what is left to do now is to express the eigenfunctions $\psi_{E}^{\sigma}(x)$ of the inverted oscillator Hamiltonian in terms of this orthonormal basis, so it is much easier to prove Eqs.~(\ref{a0}--\ref{a00}). We can accomplish this through a clever use of the Fourier transform (see Section 7.5.12 of \cite{wo79}):
\begin{equation}\label{a7}
\psi_E^{\sigma}(x) = 2^{iE/2}(2\pi)^{-1} \int_{-\infty}^{\infty} dp \, p_\sigma^{-iE - 1/2} \, e^{i(p^2/4 + xp + x^2/2)},
\end{equation}
where $p_\sigma, \ \sigma=\pm$ are defined similarly as $x_\sigma$ in Eqs.(\ref{a3}--\ref{a4}). 

Finally, the substitution of Eq.~(\ref{a7}) inside the integral of Eq.~(\ref{a0}) and the use of Eq.~(\ref{a5}) leads to the right hand side of Eq.~(\ref{a0}) and hence to the orthogonality relation. In the same way, by substituting Eq.~(\ref{a7}) into the left hand side of Eq.~(\ref{a00}) and using Eq.~(\ref{a6}) one arrives to the right hand side of Eq.~(\ref{a00}) and thus to the completeness relation.  \hfill $\square$

\newpage
\bibliographystyle{phjcp}

\end{document}